\documentclass[a4paper,12pt]{article}
\pdfoutput=1

\usepackage{undertilde}

\usepackage{amssymb} 
\usepackage{amsmath}
\usepackage{epsfig}
\usepackage{graphicx}

\makeatletter

\@addtoreset{equation}{section}
\makeatother
\def\be{\begin{equation}}
\def\ee{\end{equation}}
\def\bea{\begin{eqnarray}}
\def\eea{\end{eqnarray}}

\def\({\left(}
\def\){\right)}
\def\<{\left<}
\def\>{\right>}

\def\be{\begin{equation}}
\def\ee{\end{equation}}
\def\bea{\begin{eqnarray*}}
\def\eea{\end{eqnarray*}}
\def\ben{\begin{eqnarray}}
\def\een{\end{eqnarray}}
\def\({\left(}
\def\){\right)}
\def\<{\left<}
\def\>{\right>}
\def\!{\right|}
\def\|{\left|}

\def\[{\left[}
\def\]{\right]}

\def\+{\bar}
\def\mb{\mathbb}

\def\L{{\cal{L}}}
\def\t{\widetilde}

\def\L{{\cal{L}}}

\def\eps{{\cal{\varepsilon}}}

\def\E{{\cal{E}}}

\begin{document}

\setlength{\unitlength}{1mm}

\pagestyle{empty}
\vskip-10pt
\vskip-10pt
\hfill 
\begin{center}
\vskip 3truecm
{\Large \bf
WZW in the lightlike directions}
\vskip 2truecm
{\large \bf
Andreas Gustavsson}
\vspace{1cm} 
\begin{center} 
Physics Department, University of Seoul, Seoul 02504 KOREA
\end{center}
\vskip 0.7truecm
\begin{center}
(\tt agbrev@gmail.com)
\end{center}
\end{center}
\vskip 2truecm
{\abstract Dimensional reduction of the M5 brane on a Lorentzian manifold along a lightlike direction results in a five-dimensional gauge theory, which can be reformulated covariantly in six dimensions, where one puts the Lie derivatives along the lightlike direction of all fields to zero as constraints. Without imposing these constraints, we have a nonsupersymmetric six-dimensional gauge theory that we may expect shall have a six dimensional gauge symmetry. However this gauge symmetry has an anomaly for certain Lorentzian six-manifolds. We show that this gauge anomaly can be canceled by adding a WZW theory in the 2d space that is spanned by two lightlike directions.}

\vfill
\vskip4pt
\eject
\pagestyle{plain}

\section{Introduction} 
The nonabelian M5 brane Lagrangian is unknown. By dimensional reduction along a Killing vector field $v^M$, we obtain an ordinary five-dimensional gauge theory, but the six-dimensional covariance is lost. However, there is an alternative six-dimensionally covariant way of formulating this five-dimensional gauge theory while one has to impose as constraints that the Lie derivatives of all fields vanish along the given Killing vector field \cite{Lambert:2010wm}, \cite{Gustavsson:2018rcc}, \cite{Gustavsson:2023zny}. For Lorentzian six-manifolds there are three distinct cases depending on whether the Killing vector is spacelike $v^M v_M > 0$, timelike $v^M v_M < 0$ or lightlike $v^M v_M = 0$. 

For a spacelike Killing vector, the covariant Lagrangian is given by 
\bea
\L &=& - \frac{1}{4 g^2} F_{MN}^2 + \frac{1}{4} \eps^{MNPRST} \omega_{MNP}(A) \Omega_{RS} U_T\cr
&& - \frac{1}{2} (D_M \phi^A)^2  - \frac{R}{10} (\phi^A)^2  - \frac{g^2}{4} [\phi^A,\phi^B]^2\cr
&&  + \frac{i}{2} \bar\psi \Gamma^M D_M \psi - \frac{i}{2} \bar\psi \Gamma_M \Gamma^A [\psi,\phi^A] v^M
\eea
Here we define $\Omega_{MN} = \partial_M U_N - \partial_N U_M$ where
\bea
U_M &=& \frac{v_M}{g^2}\cr
g^2 &=& v^M v_M
\eea

For a lightlike Killing vector $v^M$, the covariant Lagrangian is given by 
\bea
\L &=& \frac{1}{2\lambda} \(G^{MN} F_{MN} + K^M K_M\) + \frac{1}{4} \eps^{MNPQRS} \omega_{MNP}(A) \Omega_{RS} U_T\cr
&& - \frac{1}{2} \(D_M \phi^A\)^2 - \frac{R}{10} \(\phi^A\)^2\cr
&& + \frac{i}{2} \bar\psi \Gamma^M D_M \psi + \frac{e}{2} \bar\psi \Gamma_M \Gamma^A [\psi,\phi^A] v^M
\eea
Here we define $\Omega_{MN} = \partial_M U_N - \partial_N U_M$ where 
\bea
U_M &=& \frac{u_M}{\lambda}\cr
\lambda &=& v^M u_M
\eea
We assume that $u^M$ is some other lightlike Killing vector, not parallel to $v^M$. For more details and the definitions of $G_{MN}$ and $K_M$ we refer to the references \cite{Gustavsson:2023zny}, \cite{Lambert:2018lgt}. 

Since we assume the existence of two Killing vectors for the lightlike dimensional reduction, while generically we assume there is just one Killing vector for the spatial dimensional reduction, it is more natural for us to consider lightlike dimensional reduction since there are then two lightlike directions in which we can put our WZW theory. We have an abelian tensor multiplet living on a single M5 brane. In particular there is a two-form gauge potential $B_{MN}$. We may form a one-form potential as the contraction $A_M = B_{MN} v^N$. We notice that the selfdual two-form has $3$ degrees of freedom, matching with the $3$ degrees of freedom in $A_M$ upon dimensional reduction. We now notice that $A_M$ if defined this way, will satisfy $A_M v^M = 0$ by the antisymmetry of $B_{MN}$. A gauge transformation $\delta B_{MN} = \partial_M \Lambda_N - \partial_N \Lambda_M$ induces a gauge variation $\delta A_M = \partial_M \lambda - \L_v \Lambda_M$ where $\lambda = \Lambda_M v^M$. This shows that by putting $\L_v \Lambda_M = 0$ and $\L_v A_M = 0$ thereby considering dimensional reduction, the gauge potential has the usual five-dimensional gauge symmetry. But this is not the only choice of $A_M$ that we can make in a six-dimensional gauge theory that will give us a five-dimensional gauge potential. Namely we can relax the condition for dimensional reduction to read $\L_v A_M = \partial_M \lambda$ where $\lambda = A_M v^M$ does not have to be zero. Still this will imply for the field strength that $\L_v F_{MN} = 0$ as one can see by noticing the identity $\L_v F_{MN} = \partial_M \L_v A_N - \partial_N \L_v A_M$. However, we still need to cut the number of components in $A_M$ since otherwise it will have 4 degrees of freedom instead of the $3$ degrees of freedom that is desired. But we can cut these number of degrees of freedom in a gauge covariant manner by imposing the constraint $F_{MN} v^N = 0$ rather than imposing a constraint such as $A_M v^M = 0$ directly on the gauge potential. A short computation\footnote{ $F_{MN} v^N = \L_v A_M - \partial_M \(v^N A_N\) = 0$ implies $\L_v F_{MN} = \partial_M \L_v A_N - \partial_N \L_v A_M = 0$.} then shows that $F_{MN} v^N = 0$ implies dimensional reduction in the sense that $\L_v F_{MN} = 0$. The upshot of this analysis is that we have a six-dimensional gauge theory with a gauge potential $A_M$ and a gauge symmetry $\delta A_M = \partial_M \lambda$ in six dimensions, on top of which we may impose $F_{MN} v^N = 0$ as a gauge covariant constraint. Dimensional reduction amounts to putting $\L_v$ on all fields to zero, up to a gauge transformation. In the nonabelian six-dimensional gauge theory, finite gauge transformations act on a scalar field and the gauge potential as
\bea
\phi^g &=& g^{-1} \phi g\cr
A_M^g &=& g^{-1} A_M g + i g^{-1} \partial_M g
\eea
If we then perform dimensional reduction in one gauge by requiring that $\L_v \phi = 0$ and $\L_v A_M = 0$, then for the gauge transformed fields, we find that
\ben
\L_v \phi^g &=& - i [\phi^g,\lambda]\cr
\L_v A_M^g &=& \partial_M \lambda - i [A_M^g,\lambda]\label{mod}
\een
where 
\bea
\lambda &=& i g^{-1} \L_v g
\eea
In many cases the difference between the dimensional reduciton of the type (\ref{mod}) and the usual dimensional reduciton $\L_v \phi = 0$ and $\L_v A_M = 0$ is inconsequential since they are gauge equivalent. There are instances when they result in different dimensional reductions, if there is a nontrivial holonomy along the direction of dimensional reduction \cite{Ishii:2007ex}. In this paper we will illustrate another instance when the gauge transformated dimensional reduction plays an important role. As we will show in this paper, it enables us to see a nonvanishing gauge anomaly as well as a nonvansihing supersymmetry anomaly from which we deduce that we shall add a certain WZW theory to cancel these anomalies.

\section{The gauge and supersymmetry anomalies}
The only potentially anomalous term in the Lagrangian is the Chern-Simons term
\bea
\L_{CS} &=& \frac{1}{4} \eps^{MNPQRS} \omega_{MNP}(A) \Omega_{QR} U_S
\eea
where the Chern-Simons three-form is defined as
\bea
\omega_{MNP}(A) &=& A_{[M} \partial_N A_{P]} - \frac{2 i}{3} A_{[M} A_N A_{P]}
\eea

Under an infinitesimal variation of the gauge potential, the Chern-Simons form transforms as 
\bea
\delta \omega_{MNP}(A) = \omega_{MNP}(A+\delta A) - \omega_{MNP}(A) = \delta A_M F_{NP} + \partial_N \(A_M \delta A_P\)
\eea
and under a finite gauge transformation it transforms as 
\bea
\delta \omega_{MNP}(A) = \omega_{MNP}(A^g) - \omega_{MNP}(A) = \partial_N \(i \partial_M g g^{-1} A_P\) + \frac{1}{3} g^{-1} \partial_M g g^{-1} \partial_N g g^{-1} \partial_P g
\eea 
By using these variations, we find the supersymmetry anomaly\footnote{Our convention for the antisymmetric tensor is such that $dx^M \wedge dx^N \wedge dx^P \wedge dx^Q \wedge dx^R \wedge dx^S = d^6 x \sqrt{-G} \eps^{MNPQRS}$ is metric-independent.} 
\bea
\delta_\eps S_{CS} &=& - \frac{1}{8} \int d^6 x \sqrt{-G} \eps^{MNPQRS} A_M i \bar\eps \Gamma_{PT} \psi v^T \Omega_{QR} \Omega_{NS}
\eea
by inserting the supersymmetry variation $\delta A_M = i \bar\eps \Gamma_{MN} \psi v^N$ and we find the gauge anomaly\footnote{We notice that $(g^{-1} dg)^3 \Omega U = d\xi$ where $\xi = (g^{-1} dg)^2 \Omega^2 - \frac{1}{2} (g^{-1} dg)^3 \Omega U$ and $U = U_M dx^M$. Being locally a total derivative means that $\int (g^{-1} dg)^3 \Omega U$ only depends on the topology. If the gauge group is $SU(2)$ and if the six-manifold has not three-cycles, then this topological invariant is zero. If the six-manifold is homeomorphic to $S^3 \times S^2 \times \mb{R}$, $\int_{S^3} (g^{-1} dg)^3 \in 24 \pi^2 \mb{Z}$ and $\int_{S^2} \Omega = 2\pi$ then we need to arrange a suitable finite range for the coordinate associated to $U$ so that $\int_{S^3} (g^{-1} dg)^3 \int_{S^2} \Omega \int U \in 2\pi \mb{Z}$.}  
\bea
\delta_g S_{CS} &=& \frac{i}{8} \int d^6 x \sqrt{-G} \eps^{MNPQRS} \partial_M g g^{-1} A_N \Omega_{PQ} \Omega_{RS}
\eea
after an integration by parts. In this paper we are assuming that $\Omega_{MN}$ satisfies the Bianchi identity $\partial_{[M} \Omega_{NP]} = 0$ everywhere. In the appendix we show that 
\bea
\Omega_{MN} v^N &=& 0\cr
\Omega_{MN} u^N &=& 0
\eea
We have a projection operator onto the lightlike directions
\bea
P_M^N &=& \frac{1}{\lambda} \(u_M v^M + v_M u^M\)
\eea
We also define
\bea
\E^{MN} &=& \frac{1}{\lambda} \(u^M v^N - u^N v^M\)\cr
\E_{MN} &=& \frac{1}{\lambda} \(u_M v_N - u_N v_M\)
\eea
These antisymmetric tensors have the property $\E^{MN} \E_{NP} = P^M_P$ which is the identity operator in the 2d subspace spanned by $u^M$ and $v^N$. We have the duality relation
\ben
\eps^{MNPQRS} \Omega_{PQ} \Omega_{RS} &=& \Omega \E^{MN}\label{dual}
\een
where the scalar function $\Omega$ is defined implicitly through this duality relation\footnote{That is, $
\Omega = - \frac{1}{\lambda} \eps^{MNPQRS} u_M v_N \Omega_{PQ} \Omega_{RS}$, which is well-defined since we assume that $\lambda$ is nowhere vanishing.}. We define projected gamma matrices 
\bea
\t\Gamma^M &=& P^M_N \Gamma^N\cr
\Sigma^M &=& \Gamma^M - \t\Gamma^M
\eea
which are mutually anticommuting, $\{\t\Gamma^M,\Sigma^N\} = 0$. We define $\Gamma_u = \Gamma^M u_M$ and $\Gamma_v = \Gamma^M v_M$. These satisfy $\Gamma_u \Gamma_u = 0 = \Gamma_v \Gamma_v$ and $\{\Gamma_u,\Gamma_v\} = 2 \lambda$. We have that $\Gamma_{uv} \Gamma_{uv} = \lambda^2$ where $\Gamma_{uv} = \Gamma^{MN} u_M v_N$. We notice that in Lorentzian signature $\lambda^2 > 0$ so $\frac{1}{\lambda}\Gamma_{uv}$ is a projection. The 6d Weyl projection can be expressed as
\bea
\t\Gamma^{MN} \Sigma^{PQRS} \eps &=& - \eps^{MNPQRS} \eps
\eea
In addition we will impose the Weyl projection (all three forms being equivalent to each other), 
\bea
\t\Gamma^{MN} \eps &=& s \E^{MN} \eps\cr
\Gamma_{uv} \eps &=& - s \lambda \eps\cr
\Sigma^{PQRS} \eps \Omega_{PQ} \Omega_{RS} &=& - s \Omega \eps
\eea
for a sign factor $s = \pm 1$. By using the duality relation (\ref{dual}) we can rewrite the anomalies in a more transparent way as
\bea
\delta_\eps S_{CS} &=& - \int d^6 x \sqrt{-G} \frac{i s \Omega}{8} v^M A_M i \bar\eps \psi 
\eea
and 
\bea
\delta_g S_{CS} &=& \int d^6 x \sqrt{-G} \frac{i \Omega}{8} \E^{MN} \partial_M g g^{-1} A_N
\eea

\section{The WZW theory}
To cancel these anomalous variations, we add the following WZW theory Lagrangian,
\bea
\L_{WZW} &=& - \frac{\Omega}{16} \(h^{-1}\(\nabla_M - i A_M\)h\)^2 + \frac{i \Omega}{8} \E^{MN} \nabla_M h h^{-1} A_N\cr
&& - \frac{\Omega}{24} \E^{MNP} h^{-1} \partial_M h h^{-1} \partial_N h h^{-1} \partial_P h\cr 
&& - \frac{i}{16} \bar\lambda\t\Gamma^M\nabla_M\lambda
\eea
This Lagrangian shall be integrated over all six dimensions in order to form the corresponding action. But we will assume that $\partial_M h$ lives in the 2d subspace of the lightlike directions, such that $\partial_M h = P_M^N \partial_N h$.
 
The supersymmetry variation is
\bea
h^{-1} \delta_{\eps} h &=& - \frac{1}{\Omega} \bar\eps \lambda\cr
\delta_{\eps} \lambda &=& - i \Gamma^M \eps h^{-1} \(\partial_M - i A_M\) h
\eea
Without also varying $A_M$, we get
\bea
\delta_{\eps}|_{\delta A_M = 0} \L_{WZW} &=& - \frac{1}{8\Omega} \bar\eps\lambda \nabla^M \Omega h^{-1} \partial_M h
\eea
For this to vanish we need to have that $\nabla^M \Omega \partial_M h = 0$, which is intuitively clear since $h$ lives in the lightlike directions and $\Omega$ lives in its orthogonal four-dimensional space. We also give a direct proof for this in the appendix.
Varying also $A_M$ according to $\delta_{\eps} A_M = i \bar\eps\Gamma_{MN} \psi v^N$ gives
\bea
\delta_{\eps} \L_{WZW} &=& \frac{i s \Omega}{8} \bar\eps \psi A_M v^M 
\eea

The gauge variation is
\bea
\lambda^g &=& \lambda\cr
h^g &=& g^{-1} h\cr
A^g_M &=& g^{-1} A_M g + i g^{-1} \partial_M g
\eea
That $\lambda$ is a gauge singlet may seem strange because it has to carry a gauge index by supersymmetry. But the variation $(h^g)^{-1} \delta_{\eps} h^g = h^{-1} g \delta_{\eps} (g^{-1} h) = h^{-1} \delta_{\eps} h$ is a gauge singlet, which is consistent with that $\lambda$ is a gauge singlet. It is also reflected in the action where we do not have the gauge covariant derivative $\nabla_M - i A_M$ acting on $\lambda$ but the ordinary curvature covariant derivative $\nabla_M \lambda$. Under this gauge transformation we have 
\bea
\delta_g \L_{WZW} &=& - \frac{i \Omega}{8} \E^{MN} \partial_M g g^{-1} A_N
\eea
By adding $\L_{CS} + \L_{WZW}$ we see that the anomalies precisely cancel.

\section*{Acknowledgement} This work was supported in part by NRF Grant RS-2023-00208011.

\appendix

\section{Proof of the relation $\nabla^M \Omega \partial_M h = 0$}
We will show that 
\bea
\nabla^M \Omega \partial_M h &=& 0
\eea
where we assume that $\partial_M h$ lives in the lightlike directions, that is, it is a linear combination of $u_M$ and $v_M$. Given this assumption of $h$, the condition on $\Omega$ will thus be that 
\bea
\L_u \Omega &=& 0\cr
\L_v \Omega &=& 0
\eea
That we shall have $\L_v \Omega = 0$ is clearly already assumed by the fact that we perform dimensional reduction along $v^M$. Furthermore $u^M$ is a Killing vector, so $\L_u g_{MN} = 0$, and the problem boils down to see whether 
\bea
\L_u \Omega_{PQ} &=& 0
\eea
or more explicitly, we shall prove that 
\bea
u^M \nabla_M \Omega_{PQ} + \nabla_P u^M \Omega_{MQ} + \nabla_Q u^M \Omega_{PM} &=& 0
\eea
We start by using 
\bea
u^M \nabla_M \lambda &=& 0
\eea
We then analyze each term in the Lie derivative separately,
\bea
u^M \nabla_M \Omega_{PQ} &=& u^M \nabla_M \nabla_P \(\frac{1}{\lambda} u_Q\) - u^M \nabla_M \nabla_Q \(\frac{1}{\lambda} u_P\)\cr
&=& \frac{1}{\lambda} \(u^M R_{MPQR} u^R - u^M R_{MQPR} u^R\)\cr
&& + u^M \nabla_P \nabla_M \(\frac{1}{\lambda} u_Q\) - u^M \nabla_Q \nabla_M \(\frac{1}{\lambda} u_P\)\cr
&=& \nabla_P \(\frac{1}{\lambda} u^M \nabla_M u_Q\) - \nabla_Q \(\frac{1}{\lambda} u^M \nabla_M u_P\)\cr
&& - \nabla_P u^M \nabla_M \(\frac{1}{\lambda} u_Q\) + \nabla_Q u^M \nabla_M \(\frac{1}{\lambda} u_P\)\cr
&=& \nabla_P \(- \frac{1}{\lambda} u^M \nabla_Q u_M\) - \nabla_Q \(- \frac{1}{\lambda} u^M \nabla_P u_M\)\cr
&& + \nabla^M u_P \nabla_M \(\frac{1}{\lambda} u_Q\) - \nabla^M u_Q \nabla_M \(\frac{1}{\lambda} u_P\)\cr
&=& \(u_Q \nabla^M u_P - u_P \nabla^M u_Q\) \nabla_M \frac{1}{\lambda}
\eea
and 
\bea
\nabla_P u^M \Omega_{MQ} &=& \nabla_P u^M \nabla_M \(\frac{1}{\lambda} u_Q\) - \nabla_P u^M \nabla_Q \(\frac{1}{\lambda} u_M\)\cr
&=& - u_Q \nabla^M u_P \nabla_M \frac{1}{\lambda} - u_M \nabla_P u^M \nabla_Q \frac{1}{\lambda}\cr
&=& - u_Q \nabla^M u_P \nabla_M \frac{1}{\lambda}
\eea
and similarly 
\bea
\nabla_Q u^M \Omega_{PM} &=& u_P \nabla^M u_Q \nabla_M \frac{1}{\lambda}
\eea
Adding up all three contributions, we find that all terms cancel, and hence we have proved that $\L_u \Omega_{PQ} = 0$.

\section{Proofs of $u^M \Omega_{MN} = 0$ and $v^M \Omega_{MN} = 0$}
First we show that $u^M \Omega_{MN} = 0$ as follows,
\bea
&& u^M \nabla_M\(\frac{1}{\lambda} u_N\) - u^M \nabla_N\(\frac{1}{\lambda} u_M\)\cr
&=& \frac{1}{\lambda} u^M \nabla_M u_N - \frac{1}{\lambda} u^M \nabla_N u_M - \frac{\nabla_M \lambda}{\lambda^2} u^M u_N + \frac{\nabla_N \lambda}{\lambda^2}  u^M u_M\cr
&=& - \frac{2}{\lambda} u^M \nabla_N u_M - \frac{\nabla_M \lambda}{\lambda^2} u^M u_N + \frac{\nabla_N \lambda}{\lambda^2}  u^M u_M\cr
&=& - \frac{1}{\lambda} \nabla_N (u^M u_M) - \frac{\nabla_M \lambda}{\lambda^2} u^M u_N + \frac{\nabla_N \lambda}{\lambda^2}  u^M u_M\cr
&=& 0
\eea
where we have used $\nabla_M u_N + \nabla_N u_M = 0$, $u^M u_M = 0$ and $u^M \nabla_M \lambda = 0$ respectively. Next we show that $v^M \Omega_{MN} = 0$ as follows,
\bea
&& v^M \nabla_M \(\frac{1}{\lambda} u_N\) - v^M \nabla_N \(\frac{1}{\lambda} u_M\)\cr
&=& - \frac{\nabla_M \lambda}{\lambda^2} v^M u_N + \frac{\nabla_N \lambda}{\lambda^2} v^M u_M + \frac{1}{\lambda} \(v^M \nabla_M u_N - v^M \nabla_N u_M\)\cr
&=& \frac{\nabla_N \lambda}{\lambda} - \frac{1}{\lambda} \(u_M \nabla_N v_M + v^M \nabla_N u_M\)\cr
&=& \frac{\nabla_N \lambda}{\lambda} - \frac{\nabla_N \(v^M u_M\)}{\lambda} \cr
&=& 0
\eea
where we have used $\L_v u_N = v^M \nabla_M u_N + \nabla_N v^M u_M = 0$.


\begin{thebibliography}{999}

\bibitem{Lambert:2010wm}
N.~Lambert and C.~Papageorgakis,
``Nonabelian (2,0) Tensor Multiplets and 3-algebras,''
JHEP \textbf{08} (2010), 083
[arXiv:1007.2982 [hep-th]].

\bibitem{Lambert:2018lgt}
N.~Lambert and M.~Owen,
``Non-Lorentzian Field Theories with Maximal Supersymmetry and Moduli Space Dynamics,''
JHEP \textbf{10} (2018), 133
[arXiv:1808.02948 [hep-th]].

\bibitem{Gustavsson:2018rcc}
A.~Gustavsson,
``The non-Abelian tensor multiplet,''
JHEP \textbf{07} (2018), 084
[arXiv:1804.04035 [hep-th]].


\bibitem{Gustavsson:2023zny}
A.~Gustavsson,
``Lightlike reduction of the M5 brane,''
JHEP \textbf{05} (2023), 130
[arXiv:2303.17846 [hep-th]].

\bibitem{Ishii:2007ex}
T.~Ishii, G.~Ishiki, S.~Shimasaki and A.~Tsuchiya,
``T-duality, fiber bundles and matrices,''
JHEP \textbf{05} (2007), 014
[arXiv:hep-th/0703021 [hep-th]].

\end{thebibliography}
\end{document}